\documentclass[aps,prr,reprint,superscriptaddress,]{revtex4-1}
\usepackage{graphics} 
\usepackage{graphicx} 

\usepackage{dcolumn}
\usepackage{bm}
\usepackage{amssymb,amsmath,xcolor}

\begin{document}
\title{Topological Nematic Spin Liquid on the Square-Kagome Lattice}
\author{Tristan Lugan}
\affiliation{Institut N\'eel, UPR2940, Univ. Grenoble Alpes et CNRS, Grenoble, FR-38042 France}

\author{L.D.C. Jaubert}
\affiliation{CNRS, Univ. Bordeaux, LOMA, UMR 5798, F-33405 Talence, France}

\author{Arnaud Ralko}
\affiliation{Institut N\'eel, UPR2940, Univ. Grenoble Alpes et CNRS, Grenoble, FR-38042 France}

\date{\today}

\begin{abstract} The ground state of the spin$-1/2$ kagome antiferromagnet
remains uncertain despite decades of active research. Here we step aside from
this debated question to address the ground-state nature of a related, and
potentially just as rich, system made of corner-sharing triangles: the
square-kagome lattice (SKL). Our work is motivated by the recent synthesis of a
distorted SKL compound mentioned in [Morita \& Tohyama, J.~Phys.~Soc.~Japan
\textbf{87}, 043704 (2018)]. We have studied its spin$-1/2$ $J_{1}$-$J_{2}$
phase diagram with an unrestricted Schwinger boson mean-field theory (SBMFT).
Our results agree with previous observations of a plaquette phase ($J_{2}\ll J_{1}$) and a ferrimagnet ($J_{1}\ll J_{2}$). In addition, three original
phases appear: two incommensurate orders and a topological quantum spin liquid
with weak nematicity. The topological order is characterized by fluxes on
specific gauge-invariant quantities and the phase is stable under anisotropic
perturbations relevant for experiments.  Finally, we provide dynamical
structure factors of the reported phases that could be observed in inelastic
neutron scattering.  \end{abstract}

\maketitle

\section{Introduction}
The search of novel topological phases is one of the most active field in
Condensed Matter, often accompanied by the emergence of fractionalized
excitations \cite{Laughlin83a,Wen91a,Moessner01c,Kitaev06a,Castelnovo12a}. The
characteristic absence of a local order parameter has made frustrated magnetism
a successful playground for topological phases, since frustration inherently
hinders order. Everyone knows the textbook example of a simple
antiferromagnetic triangle. But despite decades of research, this minimal brick
of frustration has not unveiled all its secrets. The ground state (GS) of the
quantum kagome antiferromagnet, made of corner-sharing triangles, is still
under debate between a gapless Dirac spin liquid
\cite{hermele08a,Iqbal11a,Lauchli16a,Liao17a,He17a,Ralko2018}, a gapped
$\mathbb{Z}_{2}$ spin liquid \cite{yan11a,Jiang12a,Depenbrock12a}, or something
else \cite{Lauchli16a,Ralko2018}. In this context, it is appealing to step away
from the cumbersome kagome problem, and to consider an alternate, and
potentially just as rich, network of this elementary brick of frustration.

\begin{figure}[h]
	\centering
	\includegraphics[width=0.4\textwidth,clip]{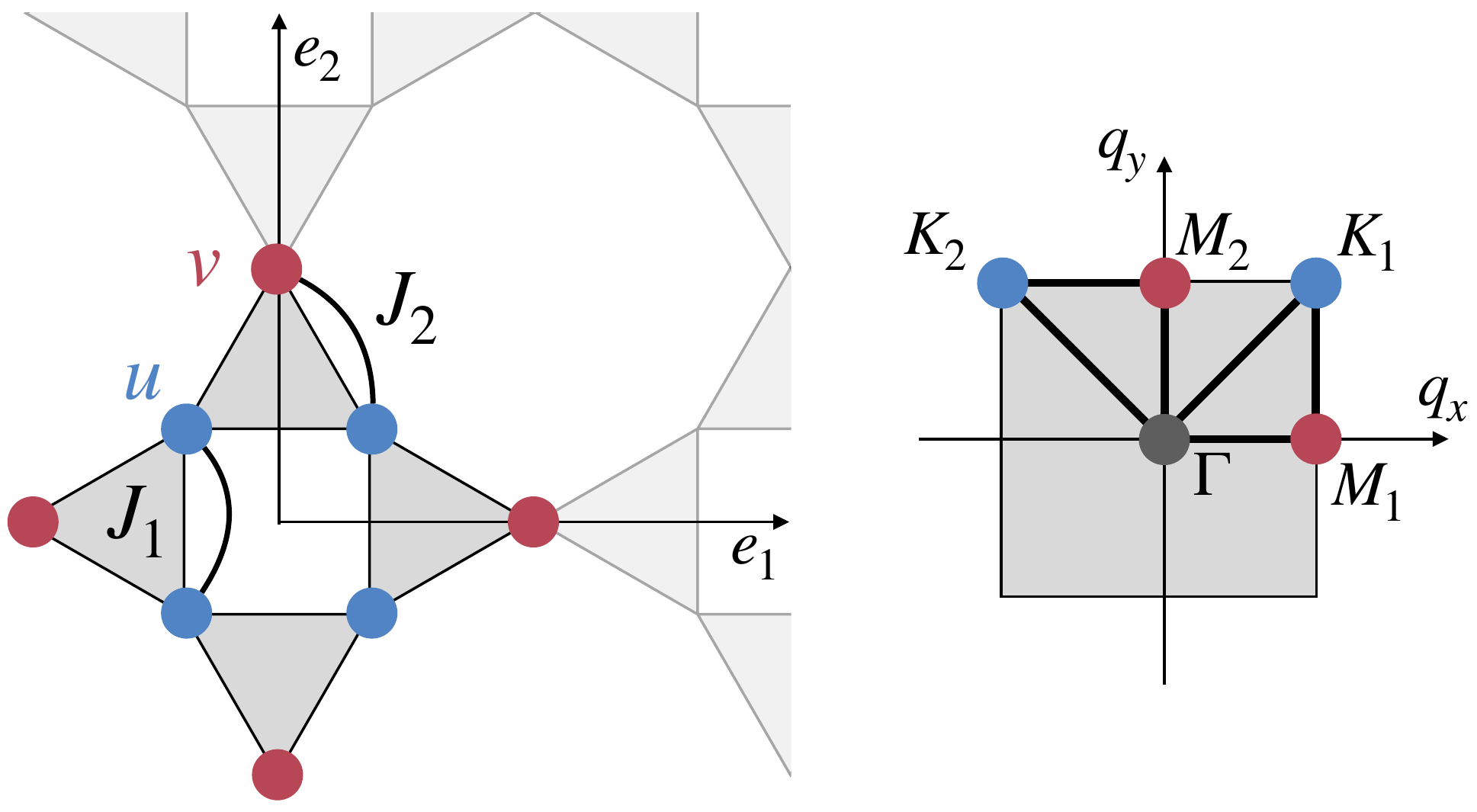}
	\caption{\small  Square-kagome lattice with its two nonequivalent sets of
sites, $u$ and $v$, respectively 4 (blue) and 2 (red) per unit cell, and the
$({\bf e}_1, {\bf e}_2)$ lattice translation vectors (left).  Spin coupling
$J_{1}$ links neighboring $u$ sites on square plaquettes while $J_{2}$ connects
$(u,v)$ sites on octagons. Two symmetry equivalent paths in the Brillouin zone
connecting high symmetry points (right).
}
	\label{fig1}
\end{figure}

The square-kagome lattice (SKL) offers such a possibility \cite{Siddharthan01a}
[Fig.~\ref{fig1}]. While the kagome lattice supports only smallest loops of 6
sites, the SKL is paved with two types of small loops of length 4 and 8. On one
hand, this asymmetry suggests a more localized quantum entanglement, and thus a
possibly more tractable treatment. On the other hand, the SKL is also a rare
and promising example where quantum resonance is \textit{not} governed by the
shortest loops \cite{Ralko2015,Ralko2018}. Adding to that the recent
announcement of a spin$-1/2$ material with distorted square-kagome geometry
\cite{Morita2018,Hasegawa18a,Fujihala2017}, and the theoretical proposal to
realise the SKL in optical lattices \cite{Glaetzle14a}, it has become a timely
question to understand the quantum GS of this model, moving beyond previous
works on exact diagonalization (ED)
\cite{Richter2009,Nakano2013,Rousochatzakis2013,Morita2018} and perturbative
methods \cite{Siddharthan01a,Tomczak03a,Ralko2015}.

\begin{figure}[h!]
\centering
\includegraphics[width=0.45\textwidth,clip]{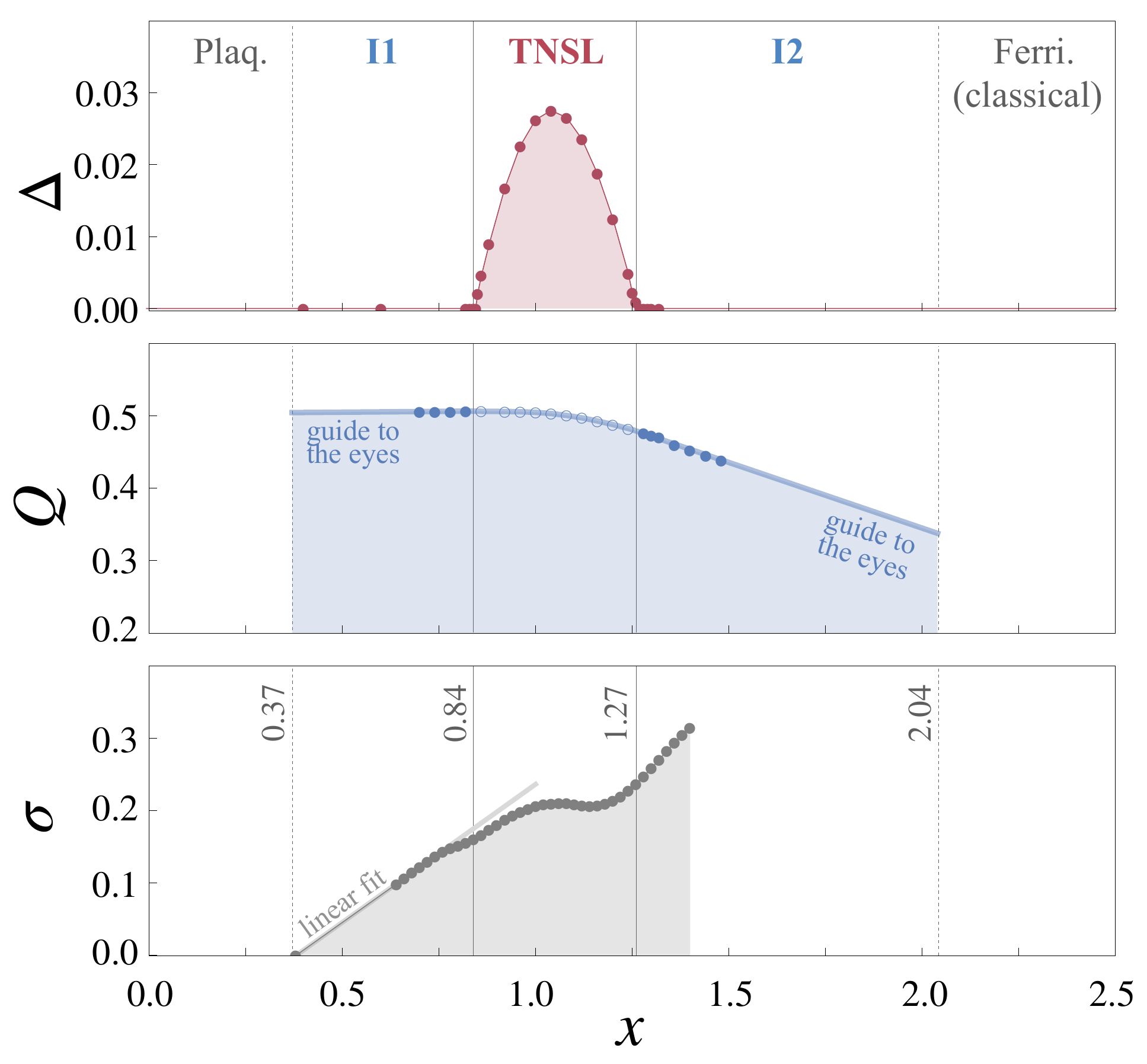}
\caption{\label{fig2}
Phase diagram of the $J_{1}$-$J_{2}$ spin$-1/2$ model on the SKL as seen
through the energy gap $\Delta$ (upper panel), the wavevector ${\bf Q} = (0,Q)$
or $(Q,0)$ minimizing the dispersion relation (middle panel) and the {\it band}
flatness $\sigma$ (lower panel) defined as the standard deviation of the energy
for a given {\it band} index.
The TNSL in the central region is characterized by a small but finite gap that
smoothly vanishes at its boundaries indicating second order
transitions (continuous vertical lines). For this phase, $Q$ corresponds to the
lowest energy excitation (empty circles).
Two incommensurate orders, I1 and I2, reflected by continuously varying
$\mathbf{Q}$ vectors at zero energy excitation (filled circles), emerge from
the crystallization of TNSL at respectively $x\simeq 0.84$ and $x\simeq 1.27$.
They extend until  energy level crossings (located at the dashed vertical
lines) with the plaquette phase at $x\simeq 0.37$ and the ferrimagnet at $x
\simeq 2.04$ through a first order transition scenario.  This is corroborated
by $\sigma$ collapsing down to zero at the transition point between I1 and the
plaquette phase.
}
\end{figure}

In this paper, we focus on the spin-$1/2$ Heisenberg model on the SKL. The
asymmetry of the lattice suggests to treat two nonequivalent bonds $J_{1}$ and
$J_{2}$, as displayed in Fig.~\ref{fig1}.  A further anisotropy $J_{2}'$,
relevant to experiments, will be considered at the end of the paper.  The
Hamiltonian then reads 
\begin{equation}
\label{eq:ham}
	\mathcal{H} = \sum \limits_{\langle i,j \rangle} J_{ij}
\mathbf{\hat{S}}_{i} \cdot \mathbf{\hat{S}}_{j},
\end{equation}
and its zero-temperature phase diagram [Fig.~\ref{fig2}] is studied by means of
an unrestricted Schwinger-boson mean-field theory (SBMFT). SBMFT is able not
only to treat on an equal footing quantum spin liquids and magnetically ordered
phases \cite{Auerbach1994} but also to reach very large sizes, a
rare feature for a quantum approach. Defining the parameter $x=J_2/J_1 >0$, we
show that for small and large values, our SBMFT is consistent with the
plaquette and ferrimagnetic phases reported in
Ref.~\onlinecite{Rousochatzakis2013}.
Then, as $x$ closes to 1 from both sides, we report the appearance of two
gapless incommensurate orders, I1 and I2. Eventually when $x\sim 1$, they give
rise to a topological nematic spin liquid (TNSL).  Finally, to make contact
with materials, we discuss the robustness of the TNSL under further anisotropy
\cite{Morita2018} and provide inelastic neutron scattering signatures of the
reported phases.

\section{Schwinger boson mean-field theory} 
SBMFT has proven successful in a variety of models: to analytically study
antiferromagnetic GS \cite{Sachdev1992}, to classify spin liquids using
projective symmetry groups \cite{Wang2006,Wang2010,Messio2013} and even to
establish phase diagrams \cite{Cabra2011,Zhang2013,Merino2018}. The method consists in introducing two Schwinger
bosons, $\hat{b}^{\dagger}_{i\uparrow}$ and $\hat{b}^{\dagger}_{i\downarrow}$, 
on each site $i$ of the lattice through the following
spin mapping
\begin{equation}
\label{spinSB}
    \mathbf{\hat{S}}_{i} = \frac{1}{2} \sum \limits_{\alpha,\beta} \hat{b}_{i,\alpha}^{\dagger} \boldsymbol{\sigma}_{\alpha,\beta} \hat{b}_{i,\beta}
\end{equation}
where $\boldsymbol{\sigma}$ are the Pauli matrices and $\alpha,\beta = \
\uparrow,\downarrow$.  For this mapping to be correct, an on-site boson
constraint $\hat{b}^{\dagger}_{i\uparrow} \hat{b}_{i\uparrow} +
\hat{b}^{\dagger}_{i\downarrow} \hat{b}_{i\downarrow} = 2S $ has to be
fulfilled.  This constraint being very hard to implement exactly, it is thus
imposed on average  by introducing Lagrange multipliers $\{\mu_{i} \}$.
Two SU(2)-invariant bond operators \cite{Flint2009}, the singlet operator on
the oriented bond $(i \rightarrow j)$ $\hat{A}_{ij} = (\hat{b}_{i\uparrow}
\hat{b}_{j\downarrow} - \hat{b}_{i\downarrow} \hat{b}_{j\uparrow})/2$ and the
boson hopping operator $\hat{B}_{ij} = (\hat{b}^{\dagger}_{i\uparrow}
\hat{b}_{j\uparrow} + \hat{b}^{\dagger}_{i\downarrow}
\hat{b}_{j\downarrow})/2$, are introduced. They are respectively favored in
disordered and ordered phases. Following a mean-field decoupling of these
operators, Eq.~\ref{eq:ham} becomes:
\begin{eqnarray}
\label{hamMF}
\hat{\mathcal{H}} &= \sum \limits_{\langle i,j \rangle} J_{ij} ( B_{ij} \hat{B}^{\dagger}_{ij} + B^{*}_{ij} \hat{B}_{ij} - A_{ij} \hat{A}^{\dagger}_{ij} - A^{*}_{ij} \hat{A}_{ij} ) \nonumber \\
 &+ \sum \limits_{\langle i,j\rangle} J_{ij} ( |A_{ij}|^{2} - |B_{ij}|^{2} ) + \sum \limits_{i} \mu_{i} ( \hat{n}_{i} - 2S )
\end{eqnarray}
where $A_{ij} = \langle \Phi_{0} | \hat{A}_{ij} | \Phi_{0}\rangle$ and $B_{ij}
= \langle \Phi_{0} | \hat{B}_{ij} | \Phi_{0}\rangle$ are mean-field parameters
computed in  the gapped boson vacuum $| \Phi_{0}\rangle$ at $T=0$.  For a
system with $n_u$ sites per unit cell, this GS is obtained by diagonalization
of the $(2 n_u) \times (2 n_u)$ $q$-dependent Hamiltonian matrix expressed in
the Fourier space -- after performing a Choleski decomposition \cite{Toth2015,
Halimeh2016, Merino2018} -- on a Brillouin zone of linear size $l$ containing
$l\times l$ momenta ($n_u \times l \times l$ lattice sites).  If one considers
translationally invariant solutions, only one Lagrange multiplier per
sublattice is needed, $\{ \mu_{s} \}_{s=0,..,n_u-1}$, as well as $4 n_u$
mean-field parameters, as shown in Fig.~\ref{fig:nemflux} (left) for $n_u = 6$.
No difference was noted on the phase diagram when using bigger unit cells up to $n_u=24$.
We choose the convention $S=(\sqrt{3}-1)/2 < 1/2$ so that our theory naturally fulfils the sum rule $\sum_i \langle \hat{\mathbf{S}}_0 \cdot
\hat{\mathbf{S}}_i \rangle = N S (S+1)$ without the extra $3/2$ factor appearing for $S=1/2$ \cite{Messio2017,Bauer2017}.

In order to compute the value of the mean-field parameters minimizing the
energy of the system, and describing the GS for different values of
$x$, we have applied the following method. First, we start from any given
\textit{Ansatz} $\{ A_{ij} , B_{ij}\}$. Then, we search for the set of Lagrange
multipliers $\{ \mu_{s} \}$ fulfilling the constraint by using a least square
minimization. After diagonalizing the Hamiltonian, we compute the new set of
parameters $\{ A_{ij} , B_{ij} \}$ from $| \Phi_{0}\rangle$.  Finally, a new
Hamiltonian is constructed and the self-consistent procedure is repeated until
convergence.  We have used an arbitrary tolerance of at least $10^{-9}$ on the
mean-field parameters and $10^{-12}$ on the energy.
This unrestricted \cite{Cabra2011, Zhang2013} method allows to find solutions
without imposing any {\it ad-hoc} \textit{Ansatz} to be miminized, that could be
too restrictive and overlook important emergent features such as our reported
TNSL. Note that this algorithm is derivative-free in contrast to usual SBMFT
approaches and naturally deals with complex mean-field parameters, crucial in
the study of gauge-invariant flux induced topological properties.

\section{Plaquette and ferrimagnet} For small values of $x$, our GS is
composed of resonating singlets on the square plaquettes formed by the $u$
sites. At the mean-field level, $v$ spins are uncorrelated leading to a large
degeneracy which is difficult to treat within our approach. However, it has
been shown that virtual plaquette excitations induce an effective coupling
between $v$ sites, responsible for a staggered valence-bond order
\cite{Rousochatzakis2013}.  Such perturbative effects are beyond SBMFT, but can
be mimicked  by introducing a coupling $J_{p}$ between $v$ sites across each
plaquette, which allows us to overcome convergence issues induced by a large degeneracy of the mean-field solution at these parameters. Taking the limit $J_{p}\to 0$, this staggered state remains
energetically favored up to $x\simeq 0.37(1)$, point at which a first order
phase transition to the I1 incommensurate order occurs.  Note that the finite
gap induced by $J_p$ vanishes as $J_{p}\to 0$ [Fig.~\ref{fig2}].  To confirm
this boundary, we have used the fact that the plaquette phase is made of local
modes -- and thus with a flat dispersion relation -- to construct a flatness
parameter $\sigma$ as the average of standard deviations of the {\it bands}.
Close to the boundary, convergence issues due to Bose condensations and
degeneracies make it difficult to measure $\sigma$, but a linear extrapolation
[gray line in Fig.~\ref{fig2}] confirms that $\sigma$ reaches zero when $x\simeq 0.38$.

For $x\gg 1$, the GS is a Lieb ferrimagnet
\cite{Lieb1989,Rousochatzakis2013,Morita2018}. Spins on sites $u$ and $v$ are
pointing in opposite directions resulting in a finite magnetization of $M=1/3$.
Classically, this phase extends to $x=2$ (resp. 1) for Heisenberg (resp.
Ising) spins \cite{Rousochatzakis2013,Pohle2016}. Our SBMFT is unfortunately
not designed for describing configurations outside the $S_{z} = 0$ sector since
it cannot fulfill the boson constraint even on average \cite{Feldner2011}.  To
estimate the boundary between I2 and the ferrimagnetic phase, we have measured
the level crossing between the extrapolation of the I2 energy measured by SBMFT
\cite{Feldner2011} and the classical energy of the ferrimagnetic phase given by
$(4S^{2}J_{1}-8S^{2}J_{2})/6$ (per site). They cross around a coherent value of
$x\approx 2.04$, but the presence of a small intervening phase -- hinted at but
not described in Refs.~\cite{Rousochatzakis2013,Morita2018} -- just below the
ferrimagnet cannot be ruled out because of convergence difficulties at large
system sizes for $x\sim 1.5 - 2.0$.

\section{Incommensurate magnetic orders} Between the plaquette and
ferrimagnetic phases, two ordered phases with Bose-Einstein condensation of the
Schwinger bosons emerge at wave vector $\mathbf{Q}$ [Fig.~\ref{fig2}].  By
spontaneous symmetry breaking, the ordering wavevector takes the form $(0,Q)$
or $(Q,0)$.  To extract the values of $Q$ and the energy gap $\Delta$ at the
thermodynamic limit, we have considered the asymptotic mean-field {\it Ansatz}
at very large system size up to $62424$ sites, before minimizing the dispersion
relation in the continuum (see Appendix A). A gapless phase with incommensurate ${\bf Q}$
vector is obtained [Fig.~\ref{fig2}] which is not
accessible by ED with periodic boundary conditions
\cite{Rousochatzakis2013,Morita2018} because of the size limitation ($\sim 30$
sites). Note however that in principle such incommensurability could be
approached by ED considering twisted boundary conditions \cite{Sindzingre2010}.
The transition points of these two gapless incommensurate phases with the
topological state are defined by the opening of the gap at $x\simeq0.84$
and $x\simeq1.27$ respectively.


\section{Topological nematic phase} But the most remarkable outcome of this
unrestricted SBMFT analysis is certainly the presence of an extended
topological nematic spin liquid (TNSL) for $x\in [ 0.84 , 1.27 ]$, as explained
below.

\begin{figure}[t]
	\centering
	\includegraphics[width=0.45\textwidth,clip]{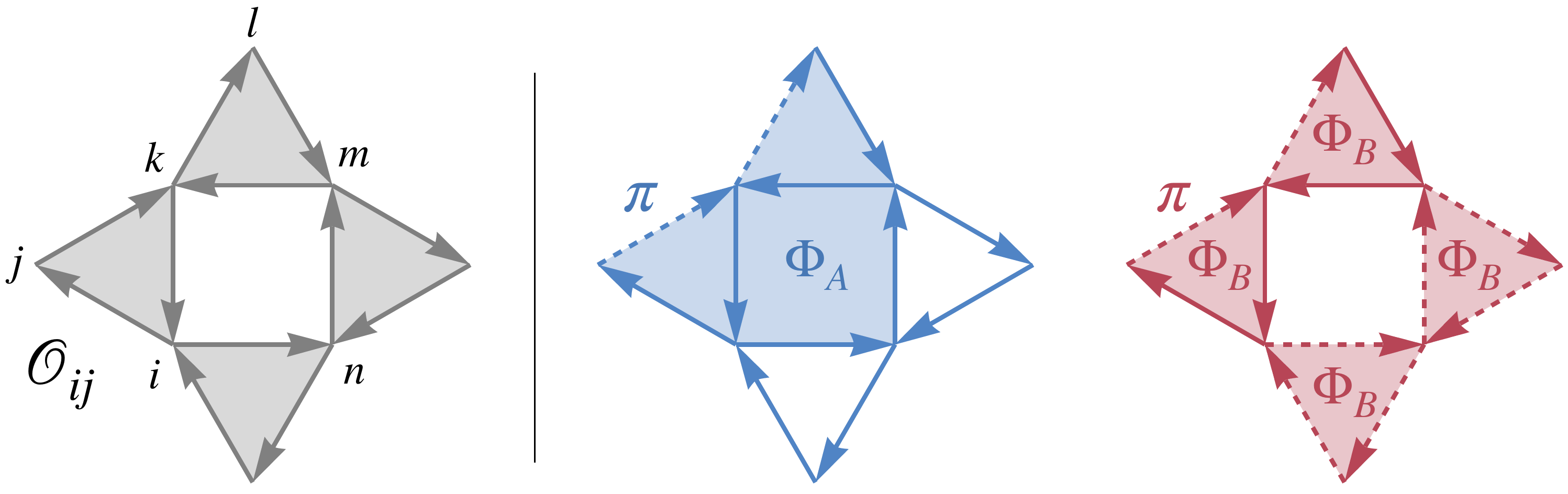}
	\caption{\label{fig:nemflux} The 24 mean-field parameters and their
orientations on the 6-site unit cell (left). Each oriented bond corresponds to
${\cal O}_{ij} = \{ A_{ij},B_{ij} \}$ as described in the text.  Dashed arrows
indicate an extra phase of $\pi$ on the corresponding mean-field parameter,
$A_{ij}$ (middle) and $B_{ij}$ (right). The blue shaded loop represents the
smallest WL with non-trivial flux $\Phi_{A}=\pi$ and the red shaded loops the 4
smallest with $\Phi_{B}=\pi$. Note that in the $A_{ij}$ WL case (middle),
$\pi$-rotation also leads to a flux of $\Phi_A = \pi$.}
\end{figure}

Within SBMFT, gauge invariant quantities called Wilson loops (WLs)
\cite{Wang2006,Tchernyshyov2006} can be constructed from  $A_{ij}$ and $B_{ij}$
parameters. We find two types of elementary plaquettes of 6 and 3 sites at
which a non-trivial gauge-invariant flux of $\pi$ emerges. On
Fig.~\ref{fig:nemflux}, this flux corresponds respectively to the phase
$\Phi_A$ of $A_{ij}(-A^{*}_{jk})A_{kl}(-A^{*}_{lm})A_{mn}(-A^{*}_{ni})$
(middle) and the phase $\Phi_B$ of $B_{ij}B_{jk}B_{ki}$ (right). Note that
$\Phi_A$ is invariant under a $\pi$-rotation of the WL [see
Fig.~\ref{fig:nemflux} (middle)]. Such non-zero fluxes
suggest the possibility of non-trivial order and help in
discriminating different phases otherwise similar by their symmetries. By defining two winding WLs around the lattice torus \cite{Bieri2016},
it is possible to evidence the topological character of the phase. A fourfold
topological degeneracy is obtained as expected for a $\mathbb{Z}_2$ spin-liquid state (see Appendix B for details). This degeneracy is enhanced to eightfold when
including the broken rotation symmetry of the nematicity (see below). It is important to note that
the physical properties and local WLs fluxes remain unchanged in each topological sector.

\begin{figure*}
\centering
\includegraphics[width=0.95\textwidth,clip]{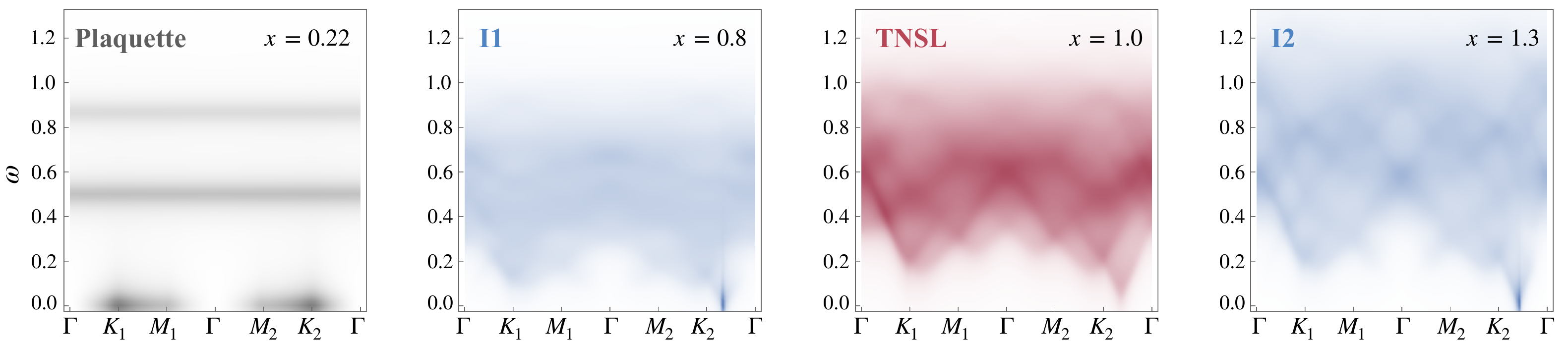}
\caption{\label{fig:SQW} Normalized dynamical structure factors of the main
different GS on the $J_{1}-J_{2}$ SKL. The path in the Brillouin zone is
composed of two $\Gamma \rightarrow K \rightarrow M \rightarrow \Gamma$ paths,
equivalent up to a $\pi/2-$rotation [see right panel of Fig.~\ref{fig1}]. The
asymmetry between these two paths in the incommensurate and nematic phases is
due to the spontaneous symmetry breaking of our {\it Ans\"atze}. The gap in the
nematic phase is not visible at this scale [see Fig.~\ref{fig2} for a zoom on
the gap].
The results for the plaquette phase have been obtained for $J_p = 10^{-3}$.
}
\end{figure*}

The gap of the TNSL reflects the presence of singlets on all bonds and indicates
an absence of long-range dipolar order. However, while being homogeneous on the
$J_1$ bonds (resonating plaquettes), singlets are weakly inhomogeneous on the
$J_{2}$ ones.  The latter on the SKL form a decorated square lattice, with
horizontal and vertical zig-zag lines. In the TNSL, the singlets on  these
horizontal lines have a different amplitudes than on the vertical ones,
breaking $\pi/2-$rotational symmetry. This nematicity can be measured by
considering a directional order parameter \cite{Zhang2013} defined as $ \Psi =
| \sum_{d=1,4} (-1)^d \langle \hat{\mathbf{S}}_{\rm v} \cdot
\hat{\mathbf{S}}_{\textrm{u}_{d}} \rangle |, \label{eq:psi4} $ where $\langle
... \rangle$ averages over all $J_2$ bonds connecting a $v$ site and one of its
four $u_{d}$ nearest-neighbors. $\Psi$ takes a small but non-zero value; {\it
e.g.} $\Psi=0.0014$ at $x=1$.

However, since the mean-field decoupling prevents access to singlet-singlet
correlations, it is relevant to compare the TNSL with other possible scenarios.
First, the weak nematicity of the TNSL motivates to consider a similar
topological \textit{Anstaz} but where rotational symmetry is restored, labeled
TSL. Then, making contact with previous works, a ``pinwheel'' valence-bond
crystal (PVBC) had also been proposed at $x=1$ \cite{Rousochatzakis2013}, later
challenged by a resonating 6-site loop (R6L) state \cite{Ralko2015} when going
beyond the nearest-neighbor valence bond basis. Interestingly, the SBMFT
counterparts of these 3 states are metastable solutions of our algorithm, with
excitation energies: magenta$\Delta E_{\textrm{TSL}}=0.0004$ $< \Delta
E_{\textrm{PVBC}}=0.0171 < \Delta E_{\textrm{R6L}}=0.0281$ at $x=1$. In
addition, the amplitudes of $\langle {\bf S_i} \cdot {\bf S_j} \rangle$ are
similar between $J_1$ and $J_2$ bonds for both the TNSL and TSL
\textit{Ans\"atze} going down to 0.1\% difference at $x=1$. Interestingly at the point $x\approx1.143$, the nematicity is lost and the ground state of our system thus becomes a quantum spin liquid. Those
energy differences and similarities in singlet amplitudes are not in favor of a
valence bond crystal (PVBC or R6L) but are consistent with a quantum spin
liquid with hidden nematic order, as defined in Ref.~\cite{Grover2010}.

\begin{figure}[h]
	\centering
	\includegraphics[width=0.35\textwidth,clip]{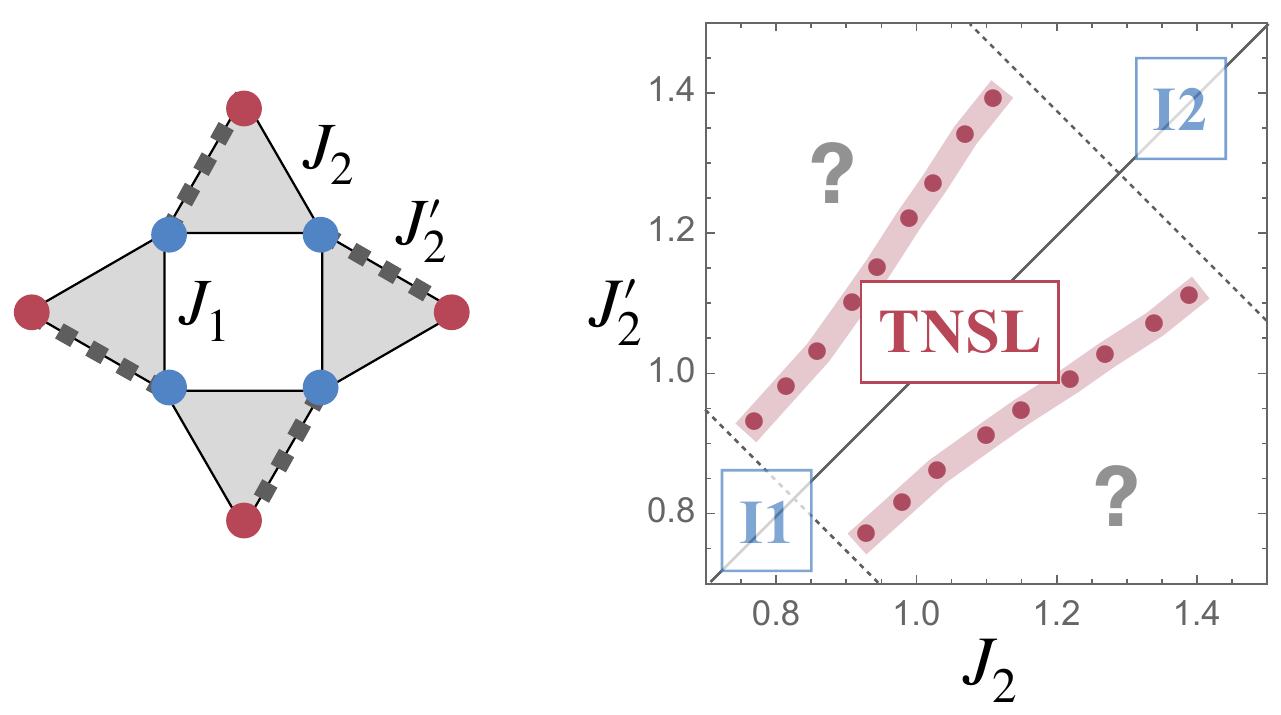}
	\caption{\label{fig5}  Distorsion induced anisotropy ($J_2,J_2'$) as could
be encountered in SKL compounds \cite{Morita2018} (left).  The isotropic $J_2 =
J_2'$ line (right) corresponds to the phase diagram of Fig.~\ref{fig2}. We have
only tested the stability of our TNSL phase against the disproportion between
$J_2$ and $J_2'$. We have tracked down the points (red circles) at which the
nematicity is lost (within our mesh precision symbolized by the thick red
lines), indicating the boundaries of the TNSL.}
\end{figure}

\section{Role of the anisotropy} As mentioned in the introduction, a
square-kagome compound has recently been announced in
Refs.~\cite{Fujihala2017,Morita2018}.  While its parametrisation is still
speculative, the authors have pointed out that treating an additional
distortion between nonequivalent $J_{2}$ bonds could be essential.  We have
then challenged our TNSL phase against such a distorsion parametrized by two
couplings $J_2$ and $J_2'$ [Fig.~\ref{fig5} left].  While the full phase
diagram of this model is beyond the scope of the present work (see
Ref.~\cite{Morita2018} for its exact-diagonalization contour), the TNSL is shown
to be stable over an extended region of the $(J_2,J_2')$ plane [Fig.~\ref{fig5}
right] making it a promising candidate for future materials. 

\section{Dynamical structure factors} A powerful experimental probe of
magnetic phases is inelastic neutron scattering, readily available by our
method via the spin-spin dynamical structure factors $S(\mathbf{q}, \omega)$
[Fig.~\ref{fig:SQW}] (see \cite{Halimeh2016,Merino2018} for technical details).
Note that we adopt the same normalization as in \cite{Halimeh2016}, setting the
maximum of $S(\mathbf{q}, \omega)$ to unity.  At small $x$,
$S(\mathbf{q},\omega)$ reveals flat bands consistent with the localized
resonating square plaquettes.  The three other phases I1, I2 and TNSL
have quite similar dynamical responses, with a clear broken symmetry indicated
by a spectral weight imbalance between the two paths in the Brillouin zone.  This
similarity, together with the smooth evolution of $Q$ across I1 and I2 in
Fig.~\ref{fig2} and the continuous gap closing, suggest that TNSL originates
from the quantum melting of the incommensurate orders.  We stress that a few
key differences (gap, intensities) would make them distinguishable in
experiments.

\section{Outlook}  The present paper opens the path for several promising
directions. We propose indeed a rare example of a topological nematic spin
liquid~\cite{Grover2010} emerging naturally from an antiferromagnetic
insulator, possibly connected to a quantum spin liquid -- see the vanishing nematic order parameter at $x\approx 1.143$. Alternative methods, such as
DMRG, would be welcome to confirm our scenario. Keeping in mind the kagome
debate, is the gap of the TNSL robust beyond mean-field ? Can we further
describe the quantum melting of the incommensurate phases ? These questions are
particularly relevant since the TNSL is stable against realistic microscopic
perturbations~\cite{Fujihala2017,Morita2018} and the SKL appears to be
realisable in optical lattices \cite{Glaetzle14a}.  Square-Kagome (artificial)
materials offer thus an exciting direction for exotic topological quantum
states.

\textit{Acknowledgements --} We thank Owen Benton, Jaime Merino,  Laura Messio,
Ioannis Rousochatzakis and Nic Shannon for fruitful discussions. The authors
thank anonymous referees for insightful questions that helped in improving the
manuscript. TL \& AR (resp. LJ) acknowledge hospitality from the LOMA in
Bordeaux (resp. the N\'eel Institute in Grenoble). This work was supported by
the ``Agence Nationale de la Recherche'' under Grant No. ANR-18-CE30-0011-01
and travel grants from the GdR MEETICC.

\appendix

\section{Reaching the thermodynamic limit}

Due to the incommensurate properties of the phases of the model, strong size
effects appear. It is thus quite delicate to properly extract the thermodynamic
limit of some quantities such as the ordering wavevector
$\mathbf{Q}_\text{soft}$ or the gap $\Delta$. Thankfully, the mean-field
parameters, that are integrated quantities (summed over the full BZ), are found
to be much more stable in function of the size. For example, in the TNSL phase,
no significant variations can be noticed for sizes larger than $l=78$ for
$n_u=6$. We have thus exploited this advantage to extract the
thermodynamic limit of the mean-field parameters instead of the physical
quantities. 

\begin{figure}[h]
	\centering
	\includegraphics[width=0.45\textwidth,clip]{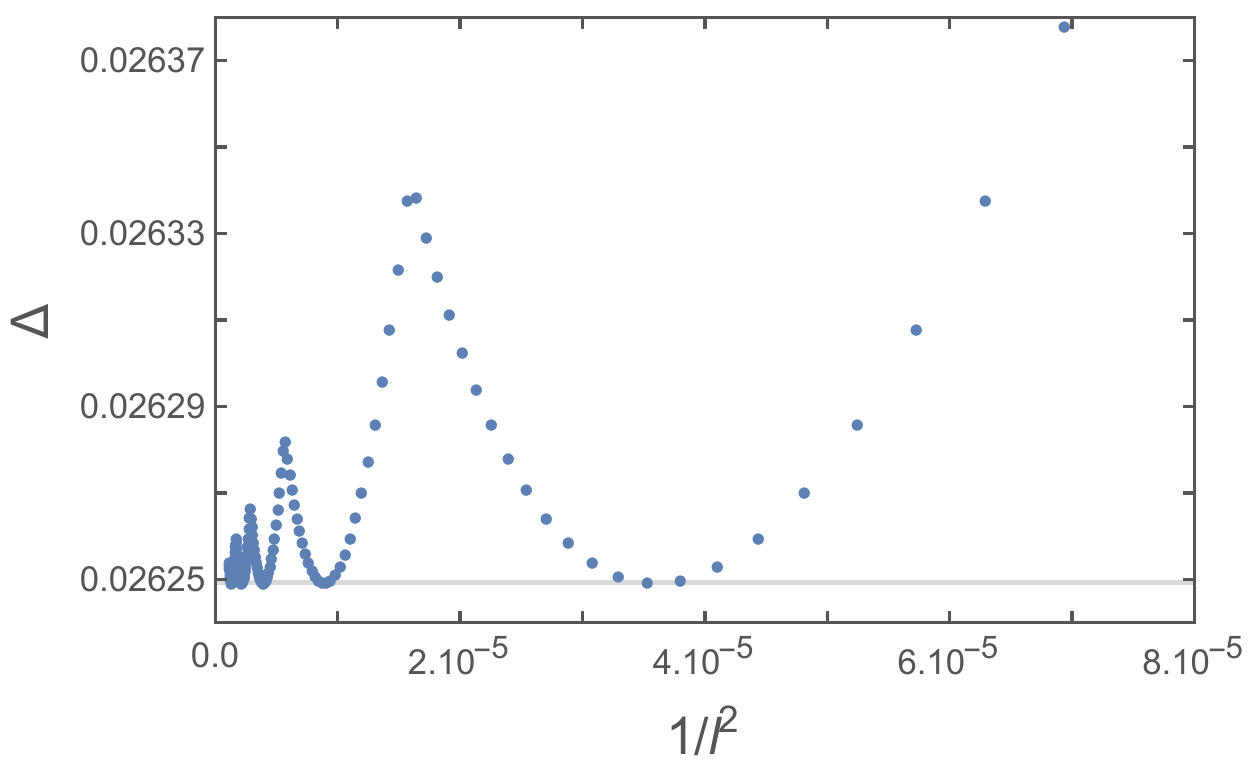}
	\caption{\label{figapp4}  Energy gap $\Delta$ versus $1/l^2$ for system
sizes ranging from $l=96$ to $l=900$ (blue dots).  The oscillations are due to
the incommensurability of the wavevector $Q_\text{soft}$. The gray line
corresponds to the thermodynamic limit of $\Delta$ computed by minimizing the
relation dispersion in the continuum for mean-field parameters extracted at the
thermodynamic limit. It also corresponds to the minimum of
the oscillations.}
\end{figure}

Minimizing the dispersion relation for these fixed Ans\"atze in the continuum of the thermodynamic limit gives access 
to $\mathbf{Q}_\text{soft}$ and $\Delta$ simultaneously, and circumventes the convergence issues.
To illustrate the efficiency of the procedure, the value of the gap for
different system sizes up to 4860000 sites is displayed on Fig.~\ref{figapp4}
(blue dots) and compared with the extracted valuemagenta in the thermodynamic limit (gray line). The oscillations reflect the incommensurability of  $\mathbf{Q}_\text{soft}$
minimizing the dispersion relation.
We note that the thermodynamic limit of $\Delta$ corresponds to the minimum of the oscillations.

This method has proven very useful to determine the
precise location of the gap opening in function of $J_{2}$. However,
numerical convergence is quite difficult beyond $l=102$ in the gapless phases.
As a result, the gap closes in a small and finite area in Fourier
space instead of a single Bragg point. Our
finite-size scalings have confirmed that this area converge to a single point
at the thermodynamic limit.

\section{Topological degeneracy}

The topological nature of the Topological Nematic Spin Liquid (TNSL) can be
evidenced by following the flux insertion procedure proposed in Bieri \textit{et al} 
\cite{Bieri2016}. The idea is that within the parton construction of quantum
spin liquids, piercing the torus by additional gauge fluxes leads to locally
indistinguishable degenerate states \cite{Bieri2016}. For lattice gauge fields
though, the flux insertion has to follow certain constraints, and a ${\mathbb
Z}_2$ quantum spin liquid only authorizes the insertion of a flux of $\pi$. 

\begin{figure}[h]
	\centering
	\includegraphics[width=0.35\textwidth,clip]{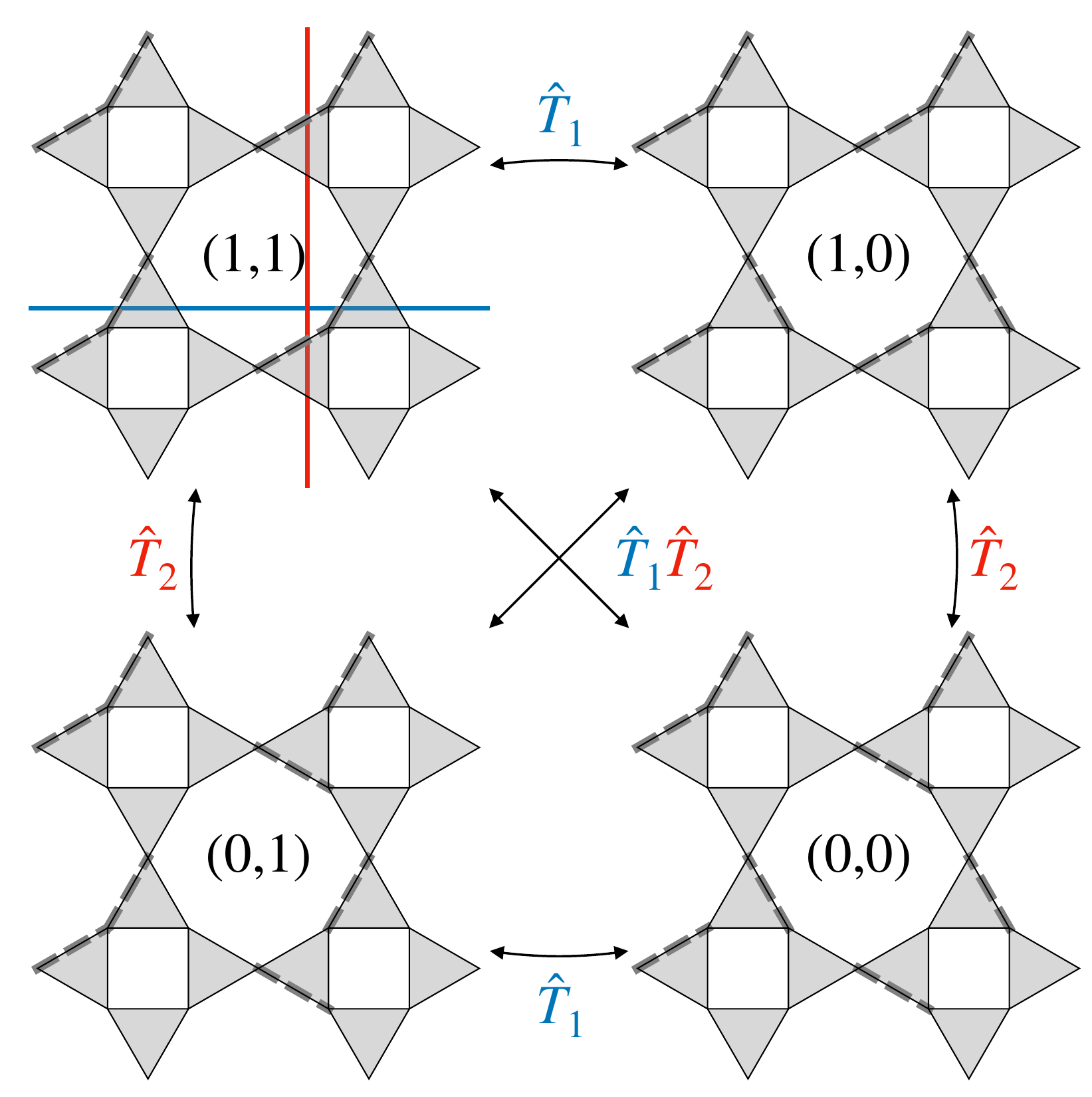}
	\caption{\label{figapp1} 
The four topological sectors obtained on the 24-site unit cell system with odd values of the system size $l$. Thick dashed bonds correspond to a phase of $-1$ on the mean-field parameters
$A_{ij}$. One can go from one gauge definition to another one by reversing all
signs of the parameters that cross the winding cut lines (red and blue) by
applying $\hat{T}_{1,2}$ operator. The topological sector is defined by the
presence or not of a $\pi$ flux along a winding WL making the complete
turn of the torus in the natural directions of the lattice (see text and Fig.~\ref{figapp2}).
}
\end{figure}

To construct these states following the constraint, one
has to define ``cut'' lines winding around the lattice in the two natural
directions as depicted in Fig.~\ref{figapp1}, and to change the sign of any
mean-field parameter crossing the line \cite{Bieri2016}.  This corresponds to
two operators, $\hat{T}_1$ and $\hat{T}_2$ verifying $\hat{T}_{1,2} {\cal
O}_{i,j} = \mp {\cal O}_{i,j}$ whether the bond $(i,j)$ crosses a cut-line or
not.
This construction is similar to the one used to determine
topological order in quantum dimer models \cite{Bieri2016}, and
is easy to apply in real space where a single cut line can be defined.
In our Fourier representation however, defining a cut line on the unit-cell
implies $l$ copies in real space. As a result, working with an enlarged
unit-cell of  $n_u=24$ sites as depicted in Fig.~\ref{figapp1}, we can generate
$l$ cut lines in each direction for a system of linear size $l$. We clearly see
that since the flux insertion can only be of $\pi$ per cut line, any even $l$
lattice size will have a trivial flux in this construction, while odd $l$ can have
non-trivial $\pi$-flux inserted.

\begin{figure}[h]
	\centering
	\includegraphics[width=0.25\textwidth,clip]{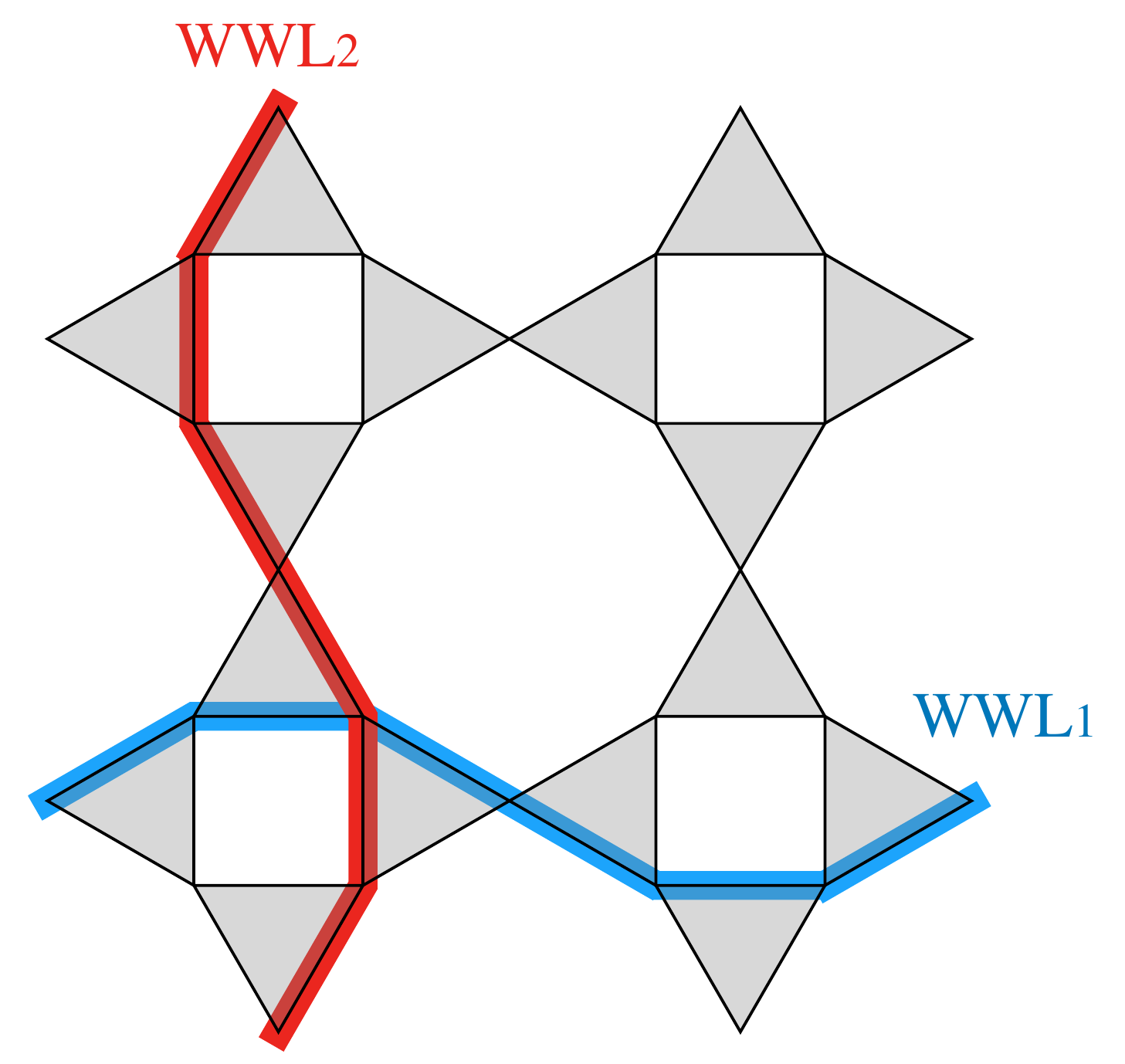}
	\caption{\label{figapp2}
Path for the two non-local winding WLs measuring the topological gauge
invariant flux of the TNSL Ans\"atze.
}
\end{figure}

Now, starting from the TNSL described in the main text and
applying operators $\hat{T}_{1,2}$ on it, 4 different Ans\"atze can be
constructed, as shown in Fig.~\ref{figapp1}. For each of them, we have
calculated through the SBMFT the energy and we have verified that the local
Wilson loops defined in the text as well as the physical parameters were
unchanged. In addition, we have defined two winding Wilson loops based on
$A_{ij}$ parameters going in the two lattice directions as depicted in
Fig.~\ref{figapp2}. The obtained fluxes are non-local and reflects the
topological nature of the phase. It is displayed in unit of $\pi$ for each
Ansatz in Fig.~\ref{figapp1}.

\begin{figure}[h]
	\centering
	\includegraphics[width=0.45\textwidth,clip]{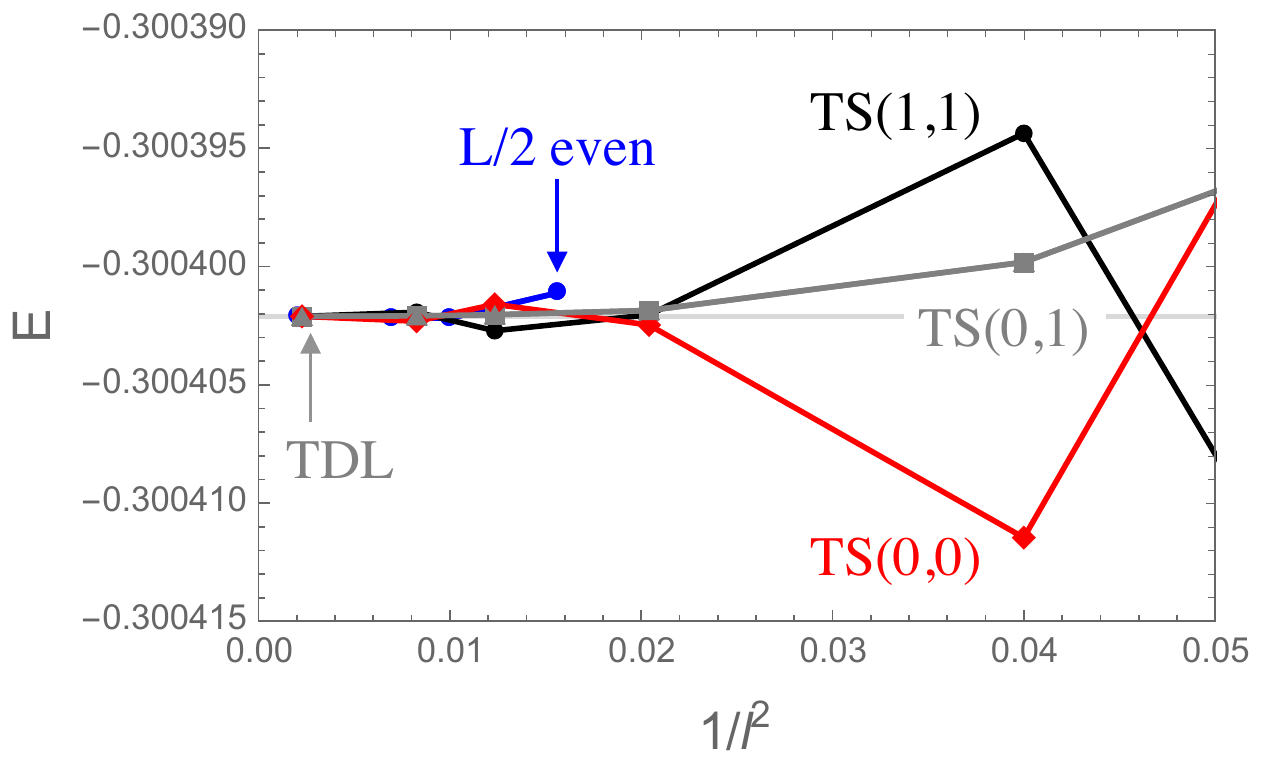}
	\caption{\label{figapp3}  
Evidence of the topological degeneracy at the thermodynamic limit. Energy of
the 4 Ans\"atze defined in Fig.~\ref{figapp1} versus $1/l^2$. The thermodynamic
limit is obtained for a 6-site unit-cell system of size $l=900$ where no
fluctuations can be observed. For $l$ even (blue points), the 4  Ans\"atze
are exactly degenerate and their winding WL fluxes are $(0,0)$. For $l$ odd,
the four topological sectors $(0,0)$ (red), $(1,1)$ (black), $(0,1)$ and
$(1,0)$ (both gray) gradually converge to the same
thermodynamic limit as $1/l^2 \to 0$.
}
\end{figure}

As mentioned previously, for a given system size $l$, the
total flux insertion of a winding WL can only be $0$ or $l \pi$. It is thus
necessarily trivial (mod $2\pi$) for even $l$; the four solutions of
Fig.~\ref{figapp1} are degenerate but do not belong to different topological
sectors. Hence, within this approach, one needs to consider odd values of $l$
to establish the topological degeneracy.
For finite and odd $l$ system size, the energies are split in three groups,
TS(0,0), TS(1,1) and TS(1,0)/(0,1), as displayed in
Fig.~\ref{figapp3}. Finite size scaling shows that these energies converge to
the same ground state energy as the one obtained for even values of $l$,
defining a thermodynamic limit as $l \to \infty$ (light
gray line). This is the fourfold topological degeneracy \cite{Misguich2012}
within the ergodicity breaking of the nematic order. This
degeneracy is enhanced to eightfold when including the rotation symmetry of the
two nematic states.

\pagebreak

\bibliographystyle{apsrev4-1} 
\bibliography{biblio} 
\end{document}